\documentclass[sigconf]{acmart}

\usepackage{booktabs}
\usepackage[table]{xcolor}
\usepackage{array}

\definecolor{NatInk}{HTML}{243447}
\definecolor{NatBlue}{HTML}{2F5D7C}
\definecolor{NatTeal}{HTML}{2A7F8F}
\definecolor{NatRule}{HTML}{B8C4CC}
\definecolor{NatWinner}{HTML}{EEF3F6}

\newcommand{\best}[1]{\textbf{\textcolor{NatInk}{#1}}}
\newcommand{\gain}[1]{\textcolor{NatTeal}{\textit{#1}}}

\definecolor{SSNavy}{HTML}{174A6E}
\definecolor{SSTeal}{HTML}{2D7F7B}
\definecolor{SSRule}{HTML}{6F8793}
\definecolor{SSHeader}{HTML}{E9F1F4}
\definecolor{SSWinner}{HTML}{E8F5F1}
\definecolor{SSGain}{HTML}{F4F8F8}

\AtBeginDocument{%
  }

\setcopyright{acmlicensed}
\copyrightyear{2026}
\acmYear{2026}
\acmConference[ACM BCB 2026]{The 17th ACM Conference on Bioinformatics, Computational Biology, and Health Informatics}{June 30--July 3, 2026}{Rende, Italy}
\acmDOI{}
\acmISBN{}
\begin{document}

\title[Stable-Shift]{Stable-Shift: Biologically Structured Prediction of Transcriptional Responses to Unseen Gene Perturbations}

\author{Sajib Acharjee Dip}
\affiliation{%
  \institution{Department of Computer Science, Virginia Tech}
  \city{Blacksburg}
  \state{VA}
  \country{USA}
}
\email{sajibacharjeedip@vt.edu}

\author{Liqing Zhang}
\authornote{Corresponding author.}
\affiliation{%
  \institution{Department of Computer Science, Virginia Tech}
  \city{Blacksburg}
  \state{VA}
  \country{USA}
}
\affiliation{%
  \institution{Fralin Biomedical Research Institute, Virginia Tech}
  \city{Roanoke}
  \state{VA}
  \country{USA}
}
\affiliation{%
  \institution{FBRI Cancer Research Center}
  \city{Washington}
  \state{DC}
  \country{USA}
}
\email{lqzhang@cs.vt.edu}

\renewcommand{\shortauthors}{Dip and Zhang}

\begin{abstract}
Predicting transcriptional responses to genetic perturbations could reduce the experimental burden of functional genomics, but extrapolation to genes that were never perturbed during training remains difficult. We present \textbf{Stable-Shift}, a structured method for estimating unseen-gene responses. Stable-Shift aggregates single-cell measurements into perturbation-level expression shifts, fits a low-rank response basis using training perturbations only, and predicts an unseen gene's coordinates in that basis from biological context. The context combines STRING interactions, network structure, control-cell expression statistics, and Gene Ontology annotations; the evaluated implementation uses graph convolution to integrate these inputs. On the supplied K562 Perturb-seq benchmark, Stable-Shift obtained 0.592 cosine similarity, compared with 0.569 for GEARS, together with higher Spearman correlation and top-gene precision among the evaluated methods. Its mean cosine similarity over five unseen-gene splits was $0.589\pm0.008$. The same ordering was observed in the supplied graph-aware, residualized, gene-space, and Norman-dataset comparisons. These results support further study of biologically structured latent-response prediction, while the lower gene-space accuracy and sensitivity to sparse graph neighborhoods limit the scope of the present conclusions.
\end{abstract}



\maketitle

\section{Introduction}

CRISPR perturbation screens coupled with single-cell RNA sequencing provide a direct way to study how gene interventions reshape cellular state \cite{dixit2016perturb,adamson2016multiplexed,dip2025llm4cell,replogle2022mapping}. Such experiments support causal gene-function discovery, target prioritization, and mechanistic analysis, but measuring every perturbation in every relevant cellular context remains infeasible. Computational models that predict the transcriptional response of an unmeasured perturbation could guide experimental design and expand the effective coverage of existing screens.

Generalization to an \emph{unseen perturbation} is substantially harder than reconstructing a held-out cell from a perturbation represented during training. The model receives no measured response for the test gene and must infer its effect from transferable gene-level context. This setting combines three difficulties: response vectors are high dimensional and noisy, perturbations can share dominant global expression programs, and the most informative relationships between genes are not necessarily visible from expression features alone.

Existing perturbation models address complementary parts of this problem. scGen and CPA learn latent representations of cell states and interventions \cite{lotfollahi2019scgen,lotfollahi2023predicting}; GEARS propagates information through a gene interaction graph \cite{roohani2024predicting}; CFM-GP predicts perturbation for unseen cell types\cite{abir2025cfm} and recent generative approaches model richer response distributions \cite{yuan2026perturbdiff}. These methods have advanced cell-level prediction, but strict extrapolation to genes excluded from training remains difficult. In particular, a feature-only model cannot exploit molecular neighborhoods at inference time, whereas a graph-only model may inherit noise and incompleteness from a single interaction resource.

We propose \textbf{Stable-Shift}, a perturbation-level method designed for unseen-gene prediction. It represents each intervention by its average expression shift from control and learns a compact response basis from training perturbations only. Stable-Shift then maps biological context available for every gene into coordinates in this basis and decodes those coordinates into a genome-wide response. The context combines a STRING interaction graph \cite{szklarczyk2019string}, Node2Vec structure \cite{grover2016node2vec}, control-cell statistics, graph summaries, and Gene Ontology (GO) annotations \cite{ashburner2000gene,gene2021gene}. This combination of a leakage-controlled response target and complementary gene-level priors defines the method; graph convolution is the encoder used in the present implementation.

Our evaluation separates train, validation, and test genes and uses identical partitions across models. Stable-Shift is compared with classical regressors, a feature-only multilayer perceptron (MLP), alternative graph architectures, and the perturbation models scGen, CPA, and GEARS. We further test sensitivity to random splits, graph-aware partitions, dominant low-rank structure, frequently responsive genes, reconstruction into the measured gene space, and a second dataset. The main contributions are:

\begin{itemize}
  \item a training-only low-rank response target that avoids using held-out perturbation measurements;
  \item a structured mapping from biological context to latent responses that combines interaction, transcriptional, topological, and functional information; and
  \item a matched evaluation that separates latent-program accuracy from gene-space reconstruction and includes several harder extrapolation tests.
\end{itemize}

\section{Related Work}

Perturbation analysis has a long history in systems biology, where interventions have been used to study signaling networks, prioritize molecular targets, and characterize sensitivity in complex biological systems \cite{kohl2010systems,mani2008systems,molinelli2013perturbation,domijan2016pettsy,caswell2000prospective,bi2024ai}. Transcriptome-scale work extends this perspective through regulatory-network inference, pathway-level response analysis, and perturbation-response gene signatures \cite{shojaie2014inferring,tegge2012pathway,schubert2018perturbation}. These studies motivate the use of biological structure, but they do not directly solve prediction for a perturbation whose response is absent from training.

Machine-learning approaches span whole-cell response prediction, multiscale function inference, gene-expression imputation, and representation learning for single-cell data \cite{ji2021machine,ma2022deep,eslami2022prediction,sailem2020kcml,chen2016gene,yuan2019deep}. Single-cell measurements can also expose genetically associated variation that is masked by bulk averages \cite{wills2013single}. This flexibility is useful for perturbation modeling, although evaluation must distinguish interpolation among observed perturbations from extrapolation to an unseen target gene.

Latent-variable models reduce the dimensionality of perturbational expression and can separate basal state from intervention effects. scGen uses latent-space arithmetic to transfer perturbation responses \cite{lotfollahi2019scgen}, while CPA learns compositional representations of perturbations and covariates \cite{lotfollahi2023predicting}. Other autoencoder and generative formulations improve flexibility but generally learn from cell states or perturbation identities observed in training \cite{gronbech2020scvae,wei2022scpregan,tang2024scperb}.

Biological graphs provide a complementary inductive bias. Molecular interactions and pathway structure can connect an unseen gene to measured perturbations, enabling neighborhood-based transfer rather than identity-based interpolation. Graph convolution, GraphSAGE, and graph attention provide different aggregation mechanisms \cite{kipf2016semi,hamilton2017inductive,velivckovic2017graph}, and GEARS applies graph propagation to perturbation prediction \cite{roohani2024predicting}. Stable-Shift uses graph aggregation within a different prediction construction: the target is a training-derived perturbation program, and the input representation combines interaction structure with control-expression and ontology information. The intended contribution is this complete response-transfer design, not a new graph convolution operator.

\begin{figure*}[t]
  \centering
  \includegraphics[width=\textwidth]{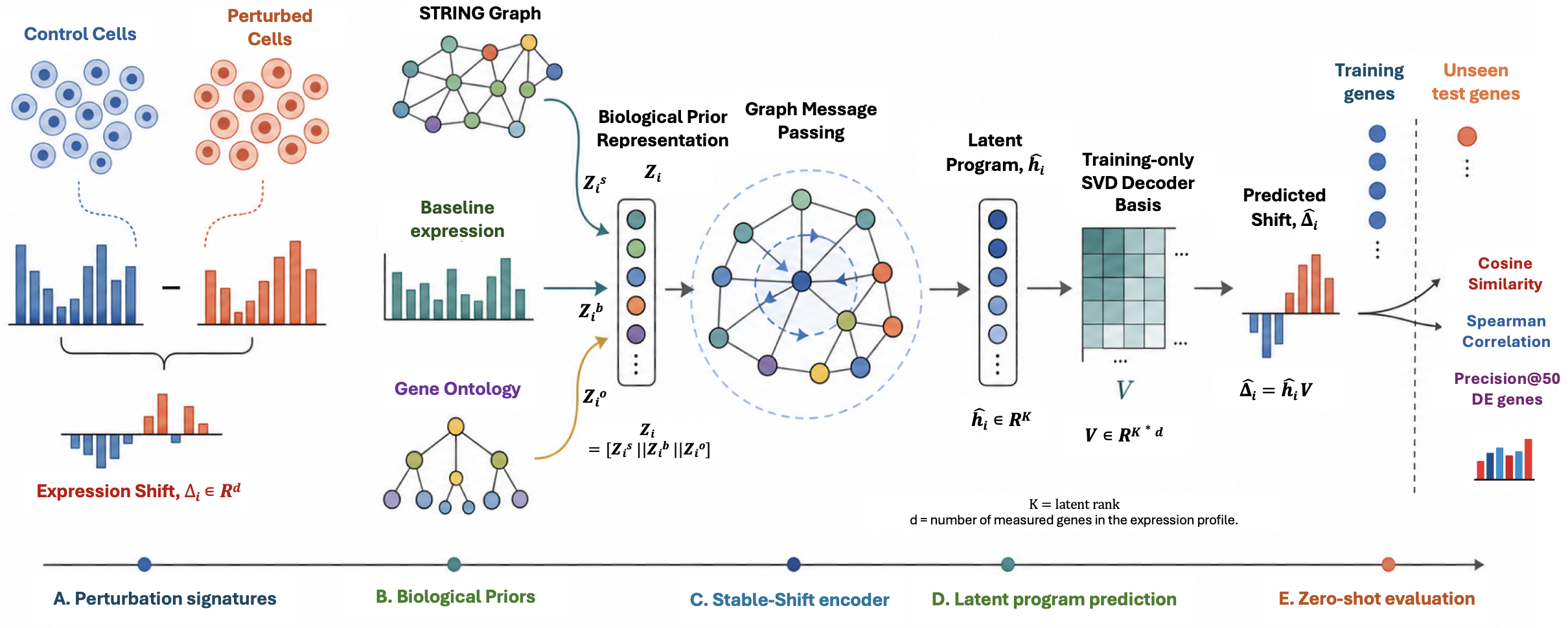}
  \caption{Stable-Shift workflow. Perturbation-level expression shifts define a training-only response basis, while complementary biological priors provide context for every gene. The Stable-Shift predictor estimates a latent response program for an unseen gene and decodes it into a genome-wide expression shift.}
  \Description{A continuous left-to-right workflow showing control and perturbed cells, expression shifts, STRING and ontology priors, Stable-Shift response prediction, latent-program decoding, and unseen-gene evaluation.}
  \label{fig:pipeline}
\end{figure*}

\section{Methods}

\subsection{Data and prediction target}

We use the K562 essential-gene Perturb-seq dataset of Replogle et al. \cite{replogle2022mapping}, containing approximately $3.1\times10^5$ cells, 8,563 measured genes, 1,832 targeted genes, and non-targeting controls. For perturbation $i$, we average the normalized expression of its cells and subtract the control mean:
\begin{equation}
  \Delta_i = x_i^{\mathrm{pert}}-x^{\mathrm{ctrl}}\in\mathbb{R}^{d}.
  \label{eq:shift}
\end{equation}
Stacking the shifts for the training perturbations gives $X_{\mathrm{train}}$. A rank-$K$ truncated singular value decomposition,
\begin{equation}
  X_{\mathrm{train}}\approx HV,
  \label{eq:svd}
\end{equation}
defines latent perturbation programs $h_i$ (rows of $H$) and a decoder $V$. Importantly, both the basis and all preprocessing statistics are fitted without validation or test responses. The latent dimension is selected by validation performance. A predicted program $\hat h_i$ is decoded as $\hat\Delta_i=\hat h_iV$.

\subsection{Biological graph and node features}

We construct a weighted gene graph from STRING protein--protein associations. Genes are nodes, retained STRING associations are edges, and edge weights are interaction-confidence scores. Self-loops are added and the adjacency is symmetrically normalized. The graph supplies a common biological context for training and unseen genes.

Each node feature vector concatenates three complementary views. First, Node2Vec embeddings summarize structural position in the interaction network. Second, control-cell statistics and graph summaries encode baseline mean expression, variance, detection frequency, degree, centrality, and local-neighborhood properties. Third, a low-dimensional GO membership embedding captures functional similarity. All response-derived transformations are fitted on the training partition.

\subsection{Stable-Shift response predictor}

Stable-Shift couples the training-derived response basis to a graph-conditioned predictor. In the implementation evaluated here, graph convolution layers integrate the feature matrix $Z$ over the normalized adjacency $\hat A$:
\begin{equation}
  H^{(\ell+1)}=\sigma\!\left(\hat A H^{(\ell)}W^{(\ell)}\right),
  \qquad H^{(0)}=Z.
  \label{eq:gcn}
\end{equation}
An MLP projection maps the resulting node representation to $\hat h_i$. Thus, graph convolution supplies the encoder, whereas the Stable-Shift prediction target is the latent response program defined from training perturbations. Training minimizes mean squared error between predicted and observed programs for training genes only:
\begin{equation}
  \mathcal{L}=|\mathcal{T}_{\mathrm{train}}|^{-1}
  \sum_{i\in\mathcal{T}_{\mathrm{train}}}\|\hat h_i-h_i\|_2^2.
  \label{eq:loss}
\end{equation}
Optimization uses Adam and validation-based early stopping. The supplement details the evaluation variants and separates settings that were previously conflated in the manuscript.

\subsection{Evaluation protocol}

Perturbation genes are partitioned into disjoint train, validation, and test sets (approximately 70/15/15). Models share the same partitions, and no test-gene response is used in fitting the SVD, features, or predictor. We report cosine similarity, Pearson and Spearman correlation, directional accuracy, mean squared error (MSE), and precision among the most strongly up- and down-regulated genes. Metrics are computed per perturbation and then averaged.

The baseline suite includes Lasso, ElasticNet, random forests, an MLP using the Stable-Shift node features, GraphSAGE, GAT, scGen, CPA, and GEARS. Cell-level baseline outputs are averaged and converted to shifts using the same control reference. The primary comparison uses the revised unified evaluation. Five independent unseen-gene splits quantify variability; separate stress tests use a graph-aware split, residualized responses, and gene-space reconstruction. Further protocol details and all secondary tables appear in the supplement.

\section{Results}

% Preamble

% Preamble

\begin{table*}[t]
\centering
\caption{\textbf{Unified unseen-gene benchmark.}
Metrics are averaged over held-out perturbations. External methods are converted to perturbation-level shifts. Bold marks Stable-Shift; underlining marks the strongest non-Stable-Shift baseline.}
\label{tab:main}
\small
\setlength{\tabcolsep}{7pt}
\renewcommand{\arraystretch}{1.18}

\begin{tabular*}{0.94\textwidth}{@{\extracolsep{\fill}}
>{\raggedright\arraybackslash}p{0.17\textwidth}
>{\raggedright\arraybackslash}p{0.23\textwidth}
cccc}
\toprule
\textcolor{NatBlue}{\textbf{Model}} &
\textcolor{NatBlue}{\textbf{Strategy}} &
\multicolumn{3}{c}{\textcolor{NatBlue}{\textbf{Predictive agreement $\uparrow$}}} &
\textcolor{NatBlue}{\textbf{Error $\downarrow$}} \\
\cmidrule(lr){3-5}\cmidrule(l){6-6}
& & \textbf{Cosine} & \textbf{Spearman} & \textbf{Prec@50} & \textbf{MSE} \\
\midrule
Random Forest & feature regression & 0.557 & 0.322 & 0.254 & 1.00 \\
scGen & single-cell latent & 0.557 & 0.318 & 0.259 & 0.96 \\
CPA & compositional latent & 0.567 & 0.326 & \underline{0.268} & 0.94 \\
GEARS & graph perturbation & \underline{0.569} & \underline{0.327} & 0.265 & -- \\
MLP & feature neural & 0.565 & 0.324 & 0.267 & \underline{0.93} \\
GraphSAGE & graph aggregation & 0.563 & 0.322 & 0.265 & 0.95 \\
\midrule
\rowcolor{NatWinner}
\best{Stable-Shift} &
\best{structured latent transfer} &
\best{0.592} &
\best{0.340} &
\best{0.277} &
\best{0.89} \\
\addlinespace[1pt]
\multicolumn{2}{l}{\gain{Gain over strongest baseline}} &
\gain{+0.023} &
\gain{+0.013} &
\gain{+0.009} &
\gain{$-0.04$} \\
\bottomrule
\end{tabular*}
\end{table*}

\subsection{Unseen-gene prediction}

Table~\ref{tab:main} reports the highest value for Stable-Shift on each available metric in this benchmark. Its cosine similarity is 0.592, an absolute difference of 0.023 from GEARS (0.569) and 0.027 from the feature-only MLP (0.565). Spearman correlation and Prec@50 show the same overall pattern, while the lower MSE is consistent with better magnitude calibration among methods reporting that metric. These are differences on one matched benchmark and should not be read as a general ranking across datasets.

The MLP receives the same gene-level features without neighborhood propagation, while the GraphSAGE variant uses graph context with a different encoder. Stable-Shift has higher reported values than both. This comparison is consistent with a benefit from the combined design, but it does not by itself assign the gain to a single component. The controlled ablations in the supplement provide more direct, although still limited, evidence: adding expression and topology statistics improves the graph-only variant, and adding GO features improves it further in that run.

\subsection{Robustness and biological fidelity}

Across five independent unseen-gene splits, Stable-Shift obtains $0.589\pm0.008$ mean cosine similarity, compared with $0.566\pm0.010$ for CPA and $0.555\pm0.012$ for random forests. A paired two-sided test over the five matched partitions gives $p<0.01$ for Stable-Shift versus CPA. Because only five partitions are available, this test is best treated as supporting evidence rather than a precise estimate of uncertainty. The analysis is distinct from the single revised benchmark in Table~\ref{tab:main}.

The supplied stress tests retain the same ordering. Under a graph-aware partition designed to separate related train and test perturbations, Stable-Shift reaches 0.558 cosine similarity versus 0.535 for GraphSAGE and 0.528 for CPA. After removing dominant shared response components, residual cosine is 0.285 versus 0.232 for CPA. Removing frequently responsive genes again preserves the ordering. These tests make simple leakage through shared or locally redundant programs less likely, but they cannot exclude it.

After decoding into the measured gene space, Stable-Shift reaches 0.392 cosine similarity versus 0.375 for CPA. This is the highest value in the supplied comparison, but it is substantially below the latent-space result and indicates that fine-grained reconstruction remains difficult. On the Norman dataset, Stable-Shift reaches 0.940 cosine and 0.815 Spearman, compared with 0.922 and 0.801 for the MLP. This second dataset broadens the evaluation, but it does not establish general transfer across cell types or experimental protocols.

\subsection{Differential-expression recovery}

Global similarity can remain high when a model recovers a dominant response direction but misses genes that matter for interpretation. We therefore also measure sign agreement, discrimination of the top differentially expressed genes, and precision among the largest predicted changes. In the controlled run reported in Table~\ref{tab:de}, Stable-Shift has directional accuracy of 0.619, DE AUROC of 0.784, DE AUPRC of 0.253, and Prec@50 of 0.292. The absolute differences from GraphSAGE are small for directional accuracy and AUROC, and larger for AUPRC and Prec@50. This pattern is consistent with better prioritization of large effects, although a larger evaluation would be needed to establish that interpretation.

\begin{table*}[t]
\centering
\caption{\textbf{Directional and differential-expression recovery in the controlled comparison.}
Prec@50 summarizes recovery of the strongest down-regulated genes. Bold marks Stable-Shift; underlining marks the strongest non-Stable-Shift baseline.}
\label{tab:de}
\small
\setlength{\tabcolsep}{9pt}
\renewcommand{\arraystretch}{1.18}

\begin{tabular*}{0.88\textwidth}{@{\extracolsep{\fill}}lrrrr}
\toprule
\textcolor{NatBlue}{\textbf{Model}} &
\multicolumn{4}{c}{\textcolor{NatBlue}{\textbf{Gene-level recovery $\uparrow$}}} \\
\cmidrule(l){2-5}
& \textbf{Directional accuracy} & \textbf{DE AUROC@50} & \textbf{DE AUPRC@50} & \textbf{Prec@50} \\
\midrule
ElasticNet & 0.609 & 0.776 & 0.214 & 0.262 \\
Random Forest & 0.614 & 0.776 & 0.218 & 0.274 \\
GraphSAGE & \underline{0.614} & \underline{0.779} & \underline{0.247} & \underline{0.278} \\
GAT & 0.613 & 0.779 & 0.224 & 0.270 \\
\midrule
\rowcolor{NatWinner}
\best{Stable-Shift} &
\best{0.619} &
\best{0.784} &
\best{0.253} &
\best{0.292} \\
\addlinespace[1pt]
\gain{Gain over GraphSAGE} &
\gain{+0.005} &
\gain{+0.005} &
\gain{+0.006} &
\gain{+0.014} \\
\bottomrule
\end{tabular*}
\end{table*}

The distinction between latent and gene-space metrics is important here. The predictor is optimized for latent-program MSE, while differential-expression scores are computed after decoding. Improvement on both levels indicates that the learned program retains useful gene-level information, but the lower gene-space cosine shows that decoding remains a major source of error. For practical use, latent similarity should therefore be reported together with at least one reconstructed-space and one top-gene metric.

\subsection{Qualitative behavior and failure modes}

\begin{figure*}[t]
  \centering
  \includegraphics[width=0.94\textwidth]{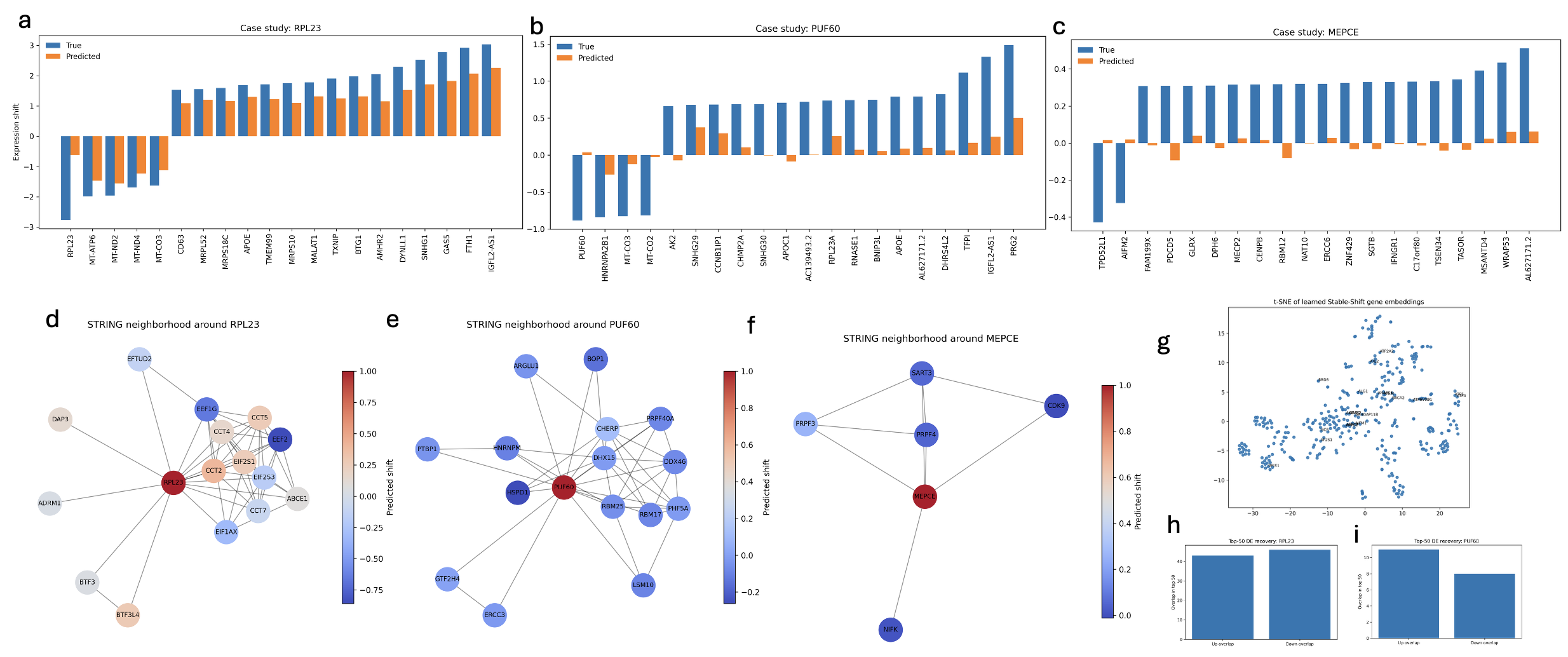}
  \caption{Qualitative analysis of Stable-Shift on unseen perturbations. Panels (a--c) compare observed and predicted shifts for selected high-performing (RPL23), median (PUF60), and low-performing (MEPCE) cases; (d--f) show their local STRING neighborhoods; (g) visualizes learned embeddings; and (h--i) summarize top-gene recovery.}
  \Description{Multi-panel figure containing expression-shift bar charts for three perturbations, three local interaction networks, a two-dimensional embedding plot, and top-gene recovery summaries.}
  \label{fig:qualitative}
\end{figure*}

Figure~\ref{fig:qualitative} examines selected high-, median-, and low-performing cases from the supplied test analysis. For \textit{RPL23}, the model reaches cosine similarity 0.969 and recovers 43 of the 50 most up-regulated genes and 46 of the 50 most down-regulated genes. The perturbation lies in a dense ribosome-associated STRING neighborhood, and predicted shifts are coherent across nearby genes. This example is compatible with the intended transfer mechanism, but a selected case cannot demonstrate that the neighborhood causes the prediction quality.

For \textit{PUF60}, cosine similarity falls to 0.335. Stable-Shift captures the direction of several large changes but underestimates their magnitude, recovering 11/50 up-regulated and 8/50 down-regulated genes. The local network remains connected but provides a less uniform response signal. For \textit{MEPCE}, cosine similarity is $-0.144$ and predicted effects are weak. Its displayed neighborhood is smaller and sparser than those of the stronger cases.

These examples reveal a useful boundary of the method: graph propagation helps most when the prior connects an unseen gene to informative measured perturbations. Sparse neighborhoods, missing regulatory edges, or perturbation-specific programs outside the low-rank basis can all cause failure. The association between neighborhood structure and these three cases is descriptive rather than causal; the graph-aware split and distance-stratified analyses in the supplement provide the more systematic tests.

\section{Discussion}

Stable-Shift should be understood as the complete mapping from biological context to a training-derived response program, rather than as a graph encoder alone. The low-rank target reduces output dimensionality and may suppress response noise; the biological representation supplies interaction, baseline-expression, topology, and ontology context for genes without measured perturbation responses. The controlled ablations are consistent with a benefit from combining these elements, although they do not exhaustively identify interactions among them.

The design also changes what generalization means. Models that receive a learned perturbation identity can interpolate among measured interventions, but that identity provides no evidence for an excluded test gene. Stable-Shift instead uses attributes available for every graph node. This makes unseen-gene inference possible, although it also ties performance to the coverage and relevance of the prior. The graph-aware split is consequently a more informative stress test than a random partition alone.

The results should be interpreted at two resolutions. Latent-program metrics measure recovery of dominant transcriptional structure and are the direct optimization target. Gene-space, differential-expression, and top-gene metrics ask whether that structure decodes into useful biological detail. Stable-Shift has higher reported values at both levels in these experiments, but the absolute drop after reconstruction remains large. A richer decoder, uncertainty estimates, or a residual pathway may be needed to preserve perturbation-specific detail.

\section{Limitations}

The present evidence is strongest for perturbation-level pseudo-bulk responses in K562. Stable-Shift does not model cell-to-cell heterogeneity, and its graph prior can inherit missing interactions and annotation bias from STRING and GO. Performance degrades in sparse graph neighborhoods and after full gene-space reconstruction. Although the Norman experiment provides a second context, broader claims about cell-type transfer require additional datasets and matched preprocessing. Finally, protein interactions are not equivalent to transcriptional regulation; integrating context-specific regulatory graphs is a promising extension.

\section*{Code Availability and Reproducibility}

The Stable-Shift source code, training and evaluation pipeline, supplementary materials, and experiment configurations are publicly available at:

\url{https://github.com/Sajib-006/PerturbGraph}

To ensure reproducibility, all experiments use fixed train/ validation/ test perturbation splits, with latent response representations learned exclusively from training data. Biological prior information, including STRING interaction networks, Gene Ontology annotations, pathway annotations, and control-cell statistics, is treated as external knowledge available before perturbation experiments and may therefore include genes appearing in the test set. The repository provides preprocessing scripts, graph construction procedures, model configurations, and evaluation code required to reproduce the reported results.

Stable-Shift is intended as a hypothesis-generation and prioritization framework for unseen perturbation prediction rather than a replacement for experimental validation. Predicted perturbation effects should be interpreted alongside supporting biological evidence and validated through downstream experimental studies when appropriate.

\section{Conclusion}

Stable-Shift combines a training-only latent response basis with complementary biological priors to predict perturbations excluded from training. Within the supplied K562 and Norman evaluations, it has higher reported values than the compared feature, graph, and perturbation baselines across several metrics. The results justify further evaluation of structured latent-response transfer, but they do not yet establish broad cross-context generalization. The remaining latent-to-gene-space gap is a concrete limitation and a useful target for future work.

\balance
\bibliographystyle{ACM-Reference-Format}
\bibliography{main}


\begin{thebibliography}{35}

%%% the Web address.

\ifx \showCODEN    \undefined \def \showCODEN     #1{\unskip}     \fi
\ifx \showISBNx    \undefined \def \showISBNx     #1{\unskip}     \fi
\ifx \showISBNxiii \undefined \def \showISBNxiii  #1{\unskip}     \fi
\ifx \showISSN     \undefined \def \showISSN      #1{\unskip}     \fi
\ifx \showLCCN     \undefined \def \showLCCN      #1{\unskip}     \fi
\ifx \shownote     \undefined \def \shownote      #1{#1}          \fi
\ifx \showarticletitle \undefined \def \showarticletitle #1{#1}   \fi
\ifx \showURL      \undefined \def \showURL       {\relax}        \fi
% invisible to TeX
\providecommand\bibfield[2]{#2}
\providecommand\bibinfo[2]{#2}
\providecommand\natexlab[1]{#1}
\providecommand\showeprint[2][]{arXiv:#2}

\bibitem[Abir et~al\mbox{.}(2025)]%
        {abir2025cfm}
\bibfield{author}{\bibinfo{person}{Abrar~Rahman Abir}, \bibinfo{person}{Sajib~Acharjee Dip}, {and} \bibinfo{person}{Liqing Zhang}.} \bibinfo{year}{2025}\natexlab{}.
\newblock \showarticletitle{CFM-GP: Unified Conditional Flow Matching to Learn Gene Perturbation Across Cell Types}.
\newblock \bibinfo{journal}{\emph{arXiv preprint arXiv:2508.08312}} (\bibinfo{year}{2025}).
\newblock


\bibitem[Adamson et~al\mbox{.}(2016)]%
        {adamson2016multiplexed}
\bibfield{author}{\bibinfo{person}{Britt Adamson}, \bibinfo{person}{Thomas~M Norman}, \bibinfo{person}{Marco Jost}, \bibinfo{person}{Min~Y Cho}, \bibinfo{person}{James~K Nu{\~n}ez}, \bibinfo{person}{Yuwen Chen}, \bibinfo{person}{Jacqueline~E Villalta}, \bibinfo{person}{Luke~A Gilbert}, \bibinfo{person}{Max~A Horlbeck}, \bibinfo{person}{Marco~Y Hein}, {et~al\mbox{.}}} \bibinfo{year}{2016}\natexlab{}.
\newblock \showarticletitle{A multiplexed single-cell CRISPR screening platform enables systematic dissection of the unfolded protein response}.
\newblock \bibinfo{journal}{\emph{Cell}} \bibinfo{volume}{167}, \bibinfo{number}{7} (\bibinfo{year}{2016}), \bibinfo{pages}{1867--1882}.
\newblock


\bibitem[Ashburner et~al\mbox{.}(2000)]%
        {ashburner2000gene}
\bibfield{author}{\bibinfo{person}{Michael Ashburner}, \bibinfo{person}{Catherine~A Ball}, \bibinfo{person}{Judith~A Blake}, \bibinfo{person}{David Botstein}, \bibinfo{person}{Heather Butler}, \bibinfo{person}{J~Michael Cherry}, \bibinfo{person}{Allan~P Davis}, \bibinfo{person}{Kara Dolinski}, \bibinfo{person}{Selina~S Dwight}, \bibinfo{person}{Janan~T Eppig}, {et~al\mbox{.}}} \bibinfo{year}{2000}\natexlab{}.
\newblock \showarticletitle{Gene ontology: tool for the unification of biology}.
\newblock \bibinfo{journal}{\emph{Nature genetics}} \bibinfo{volume}{25}, \bibinfo{number}{1} (\bibinfo{year}{2000}), \bibinfo{pages}{25--29}.
\newblock


\bibitem[Bi et~al\mbox{.}(2024)]%
        {bi2024ai}
\bibfield{author}{\bibinfo{person}{Zhenyu Bi}, \bibinfo{person}{Sajib~Acharjee Dip}, \bibinfo{person}{Daniel Hajialigol}, \bibinfo{person}{Sindhura Kommu}, \bibinfo{person}{Hanwen Liu}, \bibinfo{person}{Meng Lu}, {and} \bibinfo{person}{Xuan Wang}.} \bibinfo{year}{2024}\natexlab{}.
\newblock \showarticletitle{Ai for biomedicine in the era of large language models}.
\newblock \bibinfo{journal}{\emph{arXiv preprint arXiv:2403.15673}} (\bibinfo{year}{2024}).
\newblock


\bibitem[Caswell(2000)]%
        {caswell2000prospective}
\bibfield{author}{\bibinfo{person}{Hal Caswell}.} \bibinfo{year}{2000}\natexlab{}.
\newblock \showarticletitle{Prospective and retrospective perturbation analyses: their roles in conservation biology}.
\newblock \bibinfo{journal}{\emph{Ecology}} \bibinfo{volume}{81}, \bibinfo{number}{3} (\bibinfo{year}{2000}), \bibinfo{pages}{619--627}.
\newblock


\bibitem[Chen et~al\mbox{.}(2016)]%
        {chen2016gene}
\bibfield{author}{\bibinfo{person}{Yifei Chen}, \bibinfo{person}{Yi Li}, \bibinfo{person}{Rajiv Narayan}, \bibinfo{person}{Aravind Subramanian}, {and} \bibinfo{person}{Xiaohui Xie}.} \bibinfo{year}{2016}\natexlab{}.
\newblock \showarticletitle{Gene expression inference with deep learning}.
\newblock \bibinfo{journal}{\emph{Bioinformatics}} \bibinfo{volume}{32}, \bibinfo{number}{12} (\bibinfo{year}{2016}), \bibinfo{pages}{1832--1839}.
\newblock


\bibitem[Dip et~al\mbox{.}(2025)]%
        {dip2025llm4cell}
\bibfield{author}{\bibinfo{person}{Sajib~Acharjee Dip}, \bibinfo{person}{Adrika Zafor}, \bibinfo{person}{Bikash~Kumar Paul}, \bibinfo{person}{Uddip~Acharjee Shuvo}, \bibinfo{person}{Muhit~Islam Emon}, \bibinfo{person}{Xuan Wang}, {and} \bibinfo{person}{Liqing Zhang}.} \bibinfo{year}{2025}\natexlab{}.
\newblock \showarticletitle{LLM4Cell: A Survey of Large Language and Agentic Models for Single-Cell Biology}.
\newblock \bibinfo{journal}{\emph{arXiv preprint arXiv:2510.07793}} (\bibinfo{year}{2025}).
\newblock


\bibitem[Dixit et~al\mbox{.}(2016)]%
        {dixit2016perturb}
\bibfield{author}{\bibinfo{person}{Atray Dixit}, \bibinfo{person}{Oren Parnas}, \bibinfo{person}{Biyu Li}, \bibinfo{person}{Jenny Chen}, \bibinfo{person}{Charles~P Fulco}, \bibinfo{person}{Livnat Jerby-Arnon}, \bibinfo{person}{Nemanja~D Marjanovic}, \bibinfo{person}{Danielle Dionne}, \bibinfo{person}{Tyler Burks}, \bibinfo{person}{Raktima Raychowdhury}, {et~al\mbox{.}}} \bibinfo{year}{2016}\natexlab{}.
\newblock \showarticletitle{Perturb-Seq: dissecting molecular circuits with scalable single-cell RNA profiling of pooled genetic screens}.
\newblock \bibinfo{journal}{\emph{cell}} \bibinfo{volume}{167}, \bibinfo{number}{7} (\bibinfo{year}{2016}), \bibinfo{pages}{1853--1866}.
\newblock


\bibitem[Domijan et~al\mbox{.}(2016)]%
        {domijan2016pettsy}
\bibfield{author}{\bibinfo{person}{Mirela Domijan}, \bibinfo{person}{Paul~E Brown}, \bibinfo{person}{Boris~V Shulgin}, {and} \bibinfo{person}{David~A Rand}.} \bibinfo{year}{2016}\natexlab{}.
\newblock \showarticletitle{PeTTSy: a computational tool for perturbation analysis of complex systems biology models}.
\newblock \bibinfo{journal}{\emph{BMC bioinformatics}} \bibinfo{volume}{17}, \bibinfo{number}{1} (\bibinfo{year}{2016}), \bibinfo{pages}{124}.
\newblock


\bibitem[Eslami et~al\mbox{.}(2022)]%
        {eslami2022prediction}
\bibfield{author}{\bibinfo{person}{Mohammed Eslami}, \bibinfo{person}{Amin~Espah Borujeni}, \bibinfo{person}{Hamed Eramian}, \bibinfo{person}{Mark Weston}, \bibinfo{person}{George Zheng}, \bibinfo{person}{Joshua Urrutia}, \bibinfo{person}{Carolyn Corbet}, \bibinfo{person}{Diveena Becker}, \bibinfo{person}{Paul Maschhoff}, \bibinfo{person}{Katie Clowers}, {et~al\mbox{.}}} \bibinfo{year}{2022}\natexlab{}.
\newblock \showarticletitle{Prediction of whole-cell transcriptional response with machine learning}.
\newblock \bibinfo{journal}{\emph{Bioinformatics}} \bibinfo{volume}{38}, \bibinfo{number}{2} (\bibinfo{year}{2022}), \bibinfo{pages}{404--409}.
\newblock


\bibitem[Gr{\o}nbech et~al\mbox{.}(2020)]%
        {gronbech2020scvae}
\bibfield{author}{\bibinfo{person}{Christopher~Heje Gr{\o}nbech}, \bibinfo{person}{Maximillian~Fornitz Vording}, \bibinfo{person}{Pascal~N Timshel}, \bibinfo{person}{Casper~Kaae S{\o}nderby}, \bibinfo{person}{Tune~H Pers}, {and} \bibinfo{person}{Ole Winther}.} \bibinfo{year}{2020}\natexlab{}.
\newblock \showarticletitle{scVAE: variational auto-encoders for single-cell gene expression data}.
\newblock \bibinfo{journal}{\emph{Bioinformatics}} \bibinfo{volume}{36}, \bibinfo{number}{16} (\bibinfo{year}{2020}), \bibinfo{pages}{4415--4422}.
\newblock


\bibitem[Grover and Leskovec(2016)]%
        {grover2016node2vec}
\bibfield{author}{\bibinfo{person}{Aditya Grover} {and} \bibinfo{person}{Jure Leskovec}.} \bibinfo{year}{2016}\natexlab{}.
\newblock \showarticletitle{node2vec: Scalable feature learning for networks}. In \bibinfo{booktitle}{\emph{Proceedings of the 22nd ACM SIGKDD international conference on Knowledge discovery and data mining}}. \bibinfo{publisher}{ACM}, \bibinfo{address}{New York, NY, USA}, \bibinfo{pages}{855--864}.
\newblock


\bibitem[Hamilton et~al\mbox{.}(2017)]%
        {hamilton2017inductive}
\bibfield{author}{\bibinfo{person}{Will Hamilton}, \bibinfo{person}{Zhitao Ying}, {and} \bibinfo{person}{Jure Leskovec}.} \bibinfo{year}{2017}\natexlab{}.
\newblock \showarticletitle{Inductive representation learning on large graphs}.
\newblock \bibinfo{journal}{\emph{Advances in neural information processing systems}}  \bibinfo{volume}{30} (\bibinfo{year}{2017}), \bibinfo{pages}{1024--1034}.
\newblock


\bibitem[Ji et~al\mbox{.}(2021)]%
        {ji2021machine}
\bibfield{author}{\bibinfo{person}{Yuge Ji}, \bibinfo{person}{Mohammad Lotfollahi}, \bibinfo{person}{F~Alexander Wolf}, {and} \bibinfo{person}{Fabian~J Theis}.} \bibinfo{year}{2021}\natexlab{}.
\newblock \showarticletitle{Machine learning for perturbational single-cell omics}.
\newblock \bibinfo{journal}{\emph{Cell Systems}} \bibinfo{volume}{12}, \bibinfo{number}{6} (\bibinfo{year}{2021}), \bibinfo{pages}{522--537}.
\newblock


\bibitem[Kipf and Welling(2016)]%
        {kipf2016semi}
\bibfield{author}{\bibinfo{person}{Thomas~N Kipf} {and} \bibinfo{person}{Max Welling}.} \bibinfo{year}{2016}\natexlab{}.
\newblock \bibinfo{title}{Semi-supervised classification with graph convolutional networks}.
\newblock \bibinfo{howpublished}{arXiv preprint arXiv:1609.02907}.
\newblock


\bibitem[Kohl et~al\mbox{.}(2010)]%
        {kohl2010systems}
\bibfield{author}{\bibinfo{person}{Peter Kohl}, \bibinfo{person}{Edmund~J Crampin}, \bibinfo{person}{TA Quinn}, {and} \bibinfo{person}{Denis Noble}.} \bibinfo{year}{2010}\natexlab{}.
\newblock \showarticletitle{Systems biology: an approach}.
\newblock \bibinfo{journal}{\emph{Clinical Pharmacology \& Therapeutics}} \bibinfo{volume}{88}, \bibinfo{number}{1} (\bibinfo{year}{2010}), \bibinfo{pages}{25--33}.
\newblock


\bibitem[Lotfollahi et~al\mbox{.}(2023)]%
        {lotfollahi2023predicting}
\bibfield{author}{\bibinfo{person}{Mohammad Lotfollahi}, \bibinfo{person}{Anna Klimovskaia~Susmelj}, \bibinfo{person}{Carlo De~Donno}, \bibinfo{person}{Leon Hetzel}, \bibinfo{person}{Yuge Ji}, \bibinfo{person}{Ignacio~L Ibarra}, \bibinfo{person}{Sanjay~R Srivatsan}, \bibinfo{person}{Mohsen Naghipourfar}, \bibinfo{person}{Riza~M Daza}, \bibinfo{person}{Beth Martin}, {et~al\mbox{.}}} \bibinfo{year}{2023}\natexlab{}.
\newblock \showarticletitle{Predicting cellular responses to complex perturbations in high-throughput screens}.
\newblock \bibinfo{journal}{\emph{Molecular systems biology}} \bibinfo{volume}{19}, \bibinfo{number}{6} (\bibinfo{year}{2023}), \bibinfo{pages}{MSB202211517}.
\newblock


\bibitem[Lotfollahi et~al\mbox{.}(2019)]%
        {lotfollahi2019scgen}
\bibfield{author}{\bibinfo{person}{Mohammad Lotfollahi}, \bibinfo{person}{F~Alexander Wolf}, {and} \bibinfo{person}{Fabian~J Theis}.} \bibinfo{year}{2019}\natexlab{}.
\newblock \showarticletitle{scGen predicts single-cell perturbation responses}.
\newblock \bibinfo{journal}{\emph{Nature methods}} \bibinfo{volume}{16}, \bibinfo{number}{8} (\bibinfo{year}{2019}), \bibinfo{pages}{715--721}.
\newblock


\bibitem[Ma and Xu(2022)]%
        {ma2022deep}
\bibfield{author}{\bibinfo{person}{Qin Ma} {and} \bibinfo{person}{Dong Xu}.} \bibinfo{year}{2022}\natexlab{}.
\newblock \showarticletitle{Deep learning shapes single-cell data analysis}.
\newblock \bibinfo{journal}{\emph{Nature reviews Molecular cell biology}} \bibinfo{volume}{23}, \bibinfo{number}{5} (\bibinfo{year}{2022}), \bibinfo{pages}{303--304}.
\newblock


\bibitem[Mani et~al\mbox{.}(2008)]%
        {mani2008systems}
\bibfield{author}{\bibinfo{person}{Kartik~M Mani}, \bibinfo{person}{Celine Lefebvre}, \bibinfo{person}{Kai Wang}, \bibinfo{person}{Wei~Keat Lim}, \bibinfo{person}{Katia Basso}, \bibinfo{person}{Riccardo Dalla-Favera}, {and} \bibinfo{person}{Andrea Califano}.} \bibinfo{year}{2008}\natexlab{}.
\newblock \showarticletitle{A systems biology approach to prediction of oncogenes and molecular perturbation targets in B-cell lymphomas}.
\newblock \bibinfo{journal}{\emph{Molecular systems biology}}  \bibinfo{volume}{4} (\bibinfo{year}{2008}), \bibinfo{pages}{169}.
\newblock


\bibitem[Molinelli et~al\mbox{.}(2013)]%
        {molinelli2013perturbation}
\bibfield{author}{\bibinfo{person}{Evan~J Molinelli}, \bibinfo{person}{Anil Korkut}, \bibinfo{person}{Weiqing Wang}, \bibinfo{person}{Martin~L Miller}, \bibinfo{person}{Nicholas~P Gauthier}, \bibinfo{person}{Xiaohong Jing}, \bibinfo{person}{Poorvi Kaushik}, \bibinfo{person}{Qin He}, \bibinfo{person}{Gordon Mills}, \bibinfo{person}{David~B Solit}, {et~al\mbox{.}}} \bibinfo{year}{2013}\natexlab{}.
\newblock \showarticletitle{Perturbation biology: inferring signaling networks in cellular systems}.
\newblock \bibinfo{journal}{\emph{PLoS computational biology}} \bibinfo{volume}{9}, \bibinfo{number}{12} (\bibinfo{year}{2013}), \bibinfo{pages}{e1003290}.
\newblock


\bibitem[Replogle et~al\mbox{.}(2022)]%
        {replogle2022mapping}
\bibfield{author}{\bibinfo{person}{Joseph~M Replogle}, \bibinfo{person}{Reuben~A Saunders}, \bibinfo{person}{Angela~N Pogson}, \bibinfo{person}{Jeffrey~A Hussmann}, \bibinfo{person}{Alexander Lenail}, \bibinfo{person}{Alina Guna}, \bibinfo{person}{Lauren Mascibroda}, \bibinfo{person}{Eric~J Wagner}, \bibinfo{person}{Karen Adelman}, \bibinfo{person}{Gila Lithwick-Yanai}, {et~al\mbox{.}}} \bibinfo{year}{2022}\natexlab{}.
\newblock \showarticletitle{Mapping information-rich genotype-phenotype landscapes with genome-scale Perturb-seq}.
\newblock \bibinfo{journal}{\emph{Cell}} \bibinfo{volume}{185}, \bibinfo{number}{14} (\bibinfo{year}{2022}), \bibinfo{pages}{2559--2575}.
\newblock


\bibitem[Roohani et~al\mbox{.}(2024)]%
        {roohani2024predicting}
\bibfield{author}{\bibinfo{person}{Yusuf Roohani}, \bibinfo{person}{Kexin Huang}, {and} \bibinfo{person}{Jure Leskovec}.} \bibinfo{year}{2024}\natexlab{}.
\newblock \showarticletitle{Predicting transcriptional outcomes of novel multigene perturbations with GEARS}.
\newblock \bibinfo{journal}{\emph{Nature Biotechnology}} \bibinfo{volume}{42}, \bibinfo{number}{6} (\bibinfo{year}{2024}), \bibinfo{pages}{927--935}.
\newblock


\bibitem[Sailem et~al\mbox{.}(2020)]%
        {sailem2020kcml}
\bibfield{author}{\bibinfo{person}{Heba~Z Sailem}, \bibinfo{person}{Jens Rittscher}, {and} \bibinfo{person}{Lucas Pelkmans}.} \bibinfo{year}{2020}\natexlab{}.
\newblock \showarticletitle{KCML: a machine-learning framework for inference of multi-scale gene functions from genetic perturbation screens}.
\newblock \bibinfo{journal}{\emph{Molecular systems biology}} \bibinfo{volume}{16}, \bibinfo{number}{3} (\bibinfo{year}{2020}), \bibinfo{pages}{MSB199083}.
\newblock


\bibitem[Schubert et~al\mbox{.}(2018)]%
        {schubert2018perturbation}
\bibfield{author}{\bibinfo{person}{Michael Schubert}, \bibinfo{person}{Bertram Klinger}, \bibinfo{person}{Martina Kl{\"u}nemann}, \bibinfo{person}{Anja Sieber}, \bibinfo{person}{Florian Uhlitz}, \bibinfo{person}{Sascha Sauer}, \bibinfo{person}{Mathew~J Garnett}, \bibinfo{person}{Nils Bl{\"u}thgen}, {and} \bibinfo{person}{Julio Saez-Rodriguez}.} \bibinfo{year}{2018}\natexlab{}.
\newblock \showarticletitle{Perturbation-response genes reveal signaling footprints in cancer gene expression}.
\newblock \bibinfo{journal}{\emph{Nature communications}} \bibinfo{volume}{9}, \bibinfo{number}{1} (\bibinfo{year}{2018}), \bibinfo{pages}{20}.
\newblock


\bibitem[Shojaie et~al\mbox{.}(2014)]%
        {shojaie2014inferring}
\bibfield{author}{\bibinfo{person}{Ali Shojaie}, \bibinfo{person}{Alexandra Jauhiainen}, \bibinfo{person}{Michael Kallitsis}, {and} \bibinfo{person}{George Michailidis}.} \bibinfo{year}{2014}\natexlab{}.
\newblock \showarticletitle{Inferring regulatory networks by combining perturbation screens and steady state gene expression profiles}.
\newblock \bibinfo{journal}{\emph{PloS one}} \bibinfo{volume}{9}, \bibinfo{number}{2} (\bibinfo{year}{2014}), \bibinfo{pages}{e82393}.
\newblock


\bibitem[Szklarczyk et~al\mbox{.}(2019)]%
        {szklarczyk2019string}
\bibfield{author}{\bibinfo{person}{Damian Szklarczyk}, \bibinfo{person}{Annika~L Gable}, \bibinfo{person}{David Lyon}, \bibinfo{person}{Alexander Junge}, \bibinfo{person}{Stefan Wyder}, \bibinfo{person}{Jaime Huerta-Cepas}, \bibinfo{person}{Milan Simonovic}, \bibinfo{person}{Nadezhda~T Doncheva}, \bibinfo{person}{John~H Morris}, \bibinfo{person}{Peer Bork}, {et~al\mbox{.}}} \bibinfo{year}{2019}\natexlab{}.
\newblock \showarticletitle{STRING v11: protein--protein association networks with increased coverage, supporting functional discovery in genome-wide experimental datasets}.
\newblock \bibinfo{journal}{\emph{Nucleic acids research}} \bibinfo{volume}{47}, \bibinfo{number}{D1} (\bibinfo{year}{2019}), \bibinfo{pages}{D607--D613}.
\newblock


\bibitem[Tang et~al\mbox{.}(2024)]%
        {tang2024scperb}
\bibfield{author}{\bibinfo{person}{Zijia Tang}, \bibinfo{person}{Minghao Zhou}, \bibinfo{person}{Kai Zhang}, {and} \bibinfo{person}{Qianqian Song}.} \bibinfo{year}{2024}\natexlab{}.
\newblock \bibinfo{title}{Scperb: Predict single-cell perturbation via style transfer-based variational autoencoder}.
\newblock \bibinfo{howpublished}{Journal of Advanced Research}.
\newblock


\bibitem[Tegge et~al\mbox{.}(2012)]%
        {tegge2012pathway}
\bibfield{author}{\bibinfo{person}{Allison~N Tegge}, \bibinfo{person}{Charles~W Caldwell}, {and} \bibinfo{person}{Dong Xu}.} \bibinfo{year}{2012}\natexlab{}.
\newblock \showarticletitle{Pathway correlation profile of gene-gene co-expression for identifying pathway perturbation}.
\newblock \bibinfo{journal}{\emph{PloS one}} \bibinfo{volume}{7}, \bibinfo{number}{12} (\bibinfo{year}{2012}), \bibinfo{pages}{e52127}.
\newblock


\bibitem[{The Gene Ontology Consortium}(2021)]%
        {gene2021gene}
\bibfield{author}{\bibinfo{person}{{The Gene Ontology Consortium}}.} \bibinfo{year}{2021}\natexlab{}.
\newblock \showarticletitle{The Gene Ontology resource: enriching a GOld mine}.
\newblock \bibinfo{journal}{\emph{Nucleic acids research}} \bibinfo{volume}{49}, \bibinfo{number}{D1} (\bibinfo{year}{2021}), \bibinfo{pages}{D325--D334}.
\newblock


\bibitem[Veli{\v{c}}kovi{\'c} et~al\mbox{.}(2017)]%
        {velivckovic2017graph}
\bibfield{author}{\bibinfo{person}{Petar Veli{\v{c}}kovi{\'c}}, \bibinfo{person}{Guillem Cucurull}, \bibinfo{person}{Arantxa Casanova}, \bibinfo{person}{Adriana Romero}, \bibinfo{person}{Pietro Lio}, {and} \bibinfo{person}{Yoshua Bengio}.} \bibinfo{year}{2017}\natexlab{}.
\newblock \bibinfo{title}{Graph attention networks}.
\newblock \bibinfo{howpublished}{arXiv preprint arXiv:1710.10903}.
\newblock


\bibitem[Wei et~al\mbox{.}(2022)]%
        {wei2022scpregan}
\bibfield{author}{\bibinfo{person}{Xiajie Wei}, \bibinfo{person}{Jiayi Dong}, {and} \bibinfo{person}{Fei Wang}.} \bibinfo{year}{2022}\natexlab{}.
\newblock \showarticletitle{scPreGAN, a deep generative model for predicting the response of single-cell expression to perturbation}.
\newblock \bibinfo{journal}{\emph{Bioinformatics}} \bibinfo{volume}{38}, \bibinfo{number}{13} (\bibinfo{year}{2022}), \bibinfo{pages}{3377--3384}.
\newblock


\bibitem[Wills et~al\mbox{.}(2013)]%
        {wills2013single}
\bibfield{author}{\bibinfo{person}{Quin~F Wills}, \bibinfo{person}{Kenneth~J Livak}, \bibinfo{person}{Alex~J Tipping}, \bibinfo{person}{Tariq Enver}, \bibinfo{person}{Andrew~J Goldson}, \bibinfo{person}{Darren~W Sexton}, {and} \bibinfo{person}{Chris Holmes}.} \bibinfo{year}{2013}\natexlab{}.
\newblock \showarticletitle{Single-cell gene expression analysis reveals genetic associations masked in whole-tissue experiments}.
\newblock \bibinfo{journal}{\emph{Nature biotechnology}} \bibinfo{volume}{31}, \bibinfo{number}{8} (\bibinfo{year}{2013}), \bibinfo{pages}{748--752}.
\newblock


\bibitem[Yuan et~al\mbox{.}(2026)]%
        {yuan2026perturbdiff}
\bibfield{author}{\bibinfo{person}{Xinyu Yuan}, \bibinfo{person}{Xixian Liu}, \bibinfo{person}{Ya~Shi Zhang}, \bibinfo{person}{Zuobai Zhang}, \bibinfo{person}{Hongyu Guo}, {and} \bibinfo{person}{Jian Tang}.} \bibinfo{year}{2026}\natexlab{}.
\newblock \bibinfo{title}{PerturbDiff: Functional Diffusion for Single-Cell Perturbation Modeling}.
\newblock \bibinfo{howpublished}{arXiv preprint arXiv:2602.19685}.
\newblock


\bibitem[Yuan and Bar-Joseph(2019)]%
        {yuan2019deep}
\bibfield{author}{\bibinfo{person}{Ye Yuan} {and} \bibinfo{person}{Ziv Bar-Joseph}.} \bibinfo{year}{2019}\natexlab{}.
\newblock \showarticletitle{Deep learning for inferring gene relationships from single-cell expression data}.
\newblock \bibinfo{journal}{\emph{Proceedings of the National Academy of Sciences}} \bibinfo{volume}{116}, \bibinfo{number}{52} (\bibinfo{year}{2019}), \bibinfo{pages}{27151--27158}.
\newblock


\end{thebibliography}

\end{document}

\title{Supplementary Material for Stable-Shift: Biologically Structured Prediction of Transcriptional Responses to Unseen Gene Perturbations}
\author{Sajib Acharjee Dip}
\authornote{Corresponding author.}
\affiliation{%
  \institution{Department of Computer Science, Virginia Tech}
  \city{Blacksburg}
  \state{VA}
  \country{USA}
}
\email{sajibacharjeedip@vt.edu}

\author{Liqing Zhang}
\authornote{Corresponding author.}
\affiliation{%
  \institution{Department of Computer Science, Virginia Tech}
  \city{Blacksburg}
  \state{VA}
  \country{USA}
}
\affiliation{%
  \institution{Fralin Biomedical Research Institute, Virginia Tech}
  \city{Roanoke}
  \state{VA}
  \country{USA}
}
\affiliation{%
  \institution{FBRI Cancer Research Center}
  \city{Washington}
  \state{DC}
  \country{USA}
}
\email{lqzhang@cs.vt.edu}

\renewcommand{\shortauthors}{Dip and Zhang}
\maketitle

\section{Supplementary Evaluation Protocol}

The primary task predicts the perturbation-level expression shift of a gene whose response is absent from model training. This is stricter than holding out cells from a perturbation that the model has already observed. Perturbation genes are assigned to disjoint train, validation, and test partitions; the reported random partitions use approximately 70/15/15 percent of targeted genes. All compared methods use matched partitions and the same control reference.

The main benchmark reports the revised unified evaluation. The multi-split robustness analysis averages five separately generated unseen-gene partitions. Architecture and feature ablations were recorded in a representative controlled run and therefore should not be numerically conflated with either the revised benchmark or the five-split mean. This distinction resolves the different Stable-Shift cosine values present in the earlier manuscript.

\subsection{Perturbation signatures}

Let $x^{\mathrm{ctrl}}\in\mathbb{R}^{d}$ be the mean normalized expression vector over non-targeting control cells, and let $x_i^{\mathrm{pert}}$ be the mean over cells assigned to perturbation $i$. The intervention-level target is
\begin{equation}
  \Delta_i=x_i^{\mathrm{pert}}-x^{\mathrm{ctrl}}.
\end{equation}
This pseudo-bulk construction isolates the average shift associated with the intervention and reduces cell-level sampling noise. It does not represent cell-to-cell heterogeneity, multimodal responses, or uncertainty across cells; those limitations are carried through the present evaluation.

\subsection{Training-derived response basis}

Stacking only the training-perturbation signatures gives $X_{\mathrm{train}}\in\mathbb{R}^{n_{\mathrm{train}}\times d}$. A rank-$K$ truncated singular value decomposition defines
\begin{equation}
  X_{\mathrm{train}}\approx HV,
  \qquad H\in\mathbb{R}^{n_{\mathrm{train}}\times K},
  \quad V\in\mathbb{R}^{K\times d}.
\end{equation}
Rows of $H$ are the latent response programs used as training targets, and $V$ is the shared decoder. The rank $K$ is chosen with validation data and fixed before test evaluation. For a held-out perturbation, the predicted program $\hat h_i$ is decoded as $\hat\Delta_i=\hat h_iV$. When a latent-space test metric is reported, the observed held-out signature is projected into this fixed training basis; it is never used to refit the basis.

\subsection{Leakage controls}

The following sequence is applied independently for each partition:
\begin{enumerate}
  \item choose disjoint train, validation, and test perturbation genes;
  \item fit the response basis and response-derived preprocessing on training perturbations only;
  \item construct gene attributes that are available without observing test perturbation responses;
  \item select the latent rank and model settings using validation performance;
  \item fit the predictor with training-gene response targets; and
  \item evaluate once on the held-out genes in latent and decoded gene space.
\end{enumerate}
Graph structure, baseline control expression, and curated annotations can include test-gene nodes because these quantities are available before the test perturbation is experimentally measured. Test perturbation signatures are excluded from all fitting and selection steps.

\section{Stable-Shift Components}

\subsection{Biological graph and node representation}

Stable-Shift uses a weighted STRING association graph $\mathcal{G}=(\mathcal{V},\mathcal{E})$. If $A$ denotes its weighted adjacency matrix, self-connections and symmetric normalization give
\begin{equation}
  \begin{aligned}
    \widetilde A&=A+I,\\
    \widehat A&=\widetilde D^{-1/2}\widetilde A\widetilde D^{-1/2}.
  \end{aligned}
\end{equation}
where $\widetilde D$ is the degree matrix of $\widetilde A$. The exact confidence threshold used to retain STRING edges was not present in the supplied source and must be recovered from the experiment configuration.

Each gene representation concatenates three complementary blocks:
\begin{equation}
  z_i=\big[z_i^{s}\,\|\,z_i^{b}\,\|\,z_i^{o}\big],
\end{equation}
where $s$, $b$, and $o$ denote structural, biological-statistics, and ontology blocks. The structural block is a Node2Vec embedding of graph position. The biological block includes control-cell mean and variance, detection frequency, graph degree, centrality, and local neighborhood summaries. The functional block is a low-dimensional representation of gene--ontology membership. The Node2Vec and GO embedding dimensions were not recoverable from the manuscript and are therefore not invented here.

\subsection{Response-program predictor}

The evaluated encoder integrates node attributes with graph context using graph convolution:
\begin{equation}
  \begin{aligned}
    H^{(\ell+1)}&=\sigma\!\left(\widehat A H^{(\ell)}W^{(\ell)}\right),\\
    H^{(0)}&=Z.
  \end{aligned}
\end{equation}
A projection network maps the final node representation to $\hat h_i\in\mathbb{R}^{K}$. Parameters are learned from training perturbations by minimizing
\begin{equation}
  \mathcal L=\frac{1}{|\mathcal T_{\mathrm{train}}|}
  \sum_{i\in\mathcal T_{\mathrm{train}}}
  \|\hat h_i-h_i\|_2^2.
\end{equation}
Adam optimization and validation-based early stopping are used. Graph convolution is the encoder in this implementation; Stable-Shift denotes the full construction joining the training-derived response basis, the biological representation, the graph-conditioned predictor, and decoding back to expression space.

\section{Baseline Harmonization}

The comparison includes classical feature regressors, a feature-only multilayer perceptron, graph encoder variants, and external perturbation models. The feature-only MLP receives the same gene attributes as Stable-Shift but no graph propagation. GraphSAGE and graph-attention variants test alternative neighborhood aggregators. Random forests, Lasso, and ElasticNet provide non-graph references. External methods include scGen, CPA, and GEARS.

Methods that predict latent programs use the same training-derived decoder $V$. Methods that produce cell-level perturbed profiles are converted to the common target by averaging predicted cells for each perturbation and subtracting the same control mean. This conversion does not make the model classes identical, but it ensures that the reported metrics compare perturbation-level shifts on matched genes rather than mixing cell-level and pseudo-bulk targets.

\begin{table*}[t]
  \centering
  \caption{Role and output harmonization of the model families used in the evaluation.}
  \label{tab:harmonization}
  \begingroup
  \small
  \renewcommand{\arraystretch}{1.14}
  \arrayrulecolor{SSRule}
  \begin{tabular*}{0.92\textwidth}{@{\extracolsep{\fill}}>{\raggedright\arraybackslash}p{0.18\textwidth}>{\raggedright\arraybackslash}p{0.23\textwidth}>{\raggedright\arraybackslash}p{0.25\textwidth}>{\raggedright\arraybackslash}p{0.15\textwidth}}
    \toprule
    \rowcolor{SSHeader}
    \ssth{Model family} & \ssth{Native prediction} & \ssth{Conversion for evaluation} & \ssth{Primary role} \\
    \midrule
    Lasso, ElasticNet, random forest & latent response coordinates & decode with the shared training basis & classical feature reference \\
    MLP & latent response coordinates & decode with the shared training basis & feature-only neural reference \\
    GraphSAGE, GAT & latent response coordinates & decode with the shared training basis & encoder comparison \\
    scGen, CPA, GEARS & cell-level perturbed expression & average predicted cells and subtract the common control mean & external perturbation reference \\
    \rowcolor{SSWinner}
    \sstbest{Stable-Shift} & \sstbest{latent response coordinates} & \sstbest{decode with the shared training basis} & \sstbest{proposed method} \\
    \bottomrule
  \end{tabular*}
  \endgroup
\end{table*}

\section{Metric Definitions}

For each test perturbation $i$, metrics compare the observed and predicted expression shifts, $\Delta_i$ and $\hat\Delta_i$. Cosine similarity measures vector orientation:
\begin{equation}
  \mathrm{Cosine}_i=
  \frac{\Delta_i^{\top}\hat\Delta_i}
       {\|\Delta_i\|_2\,\|\hat\Delta_i\|_2}.
\end{equation}
Spearman correlation is the Pearson correlation between the gene-wise ranks of the two vectors and is less sensitive to global scale. Magnitude error is reported as
\begin{equation}
  \mathrm{MSE}_i=\frac{1}{d}\|\Delta_i-\hat\Delta_i\|_2^2.
\end{equation}
Directional accuracy is
\begin{equation}
  \mathrm{DirAcc}_i=\frac{1}{d}\sum_{j=1}^{d}
  \mathbb{I}\!\left[\mathrm{sign}(\Delta_{ij})=
  \mathrm{sign}(\hat\Delta_{ij})\right].
\end{equation}
For top-gene recovery, let $U_k(v)$ and $D_k(v)$ denote the indices of the $k$ largest positive and most negative entries of $v$. Then
\begin{equation}
  \mathrm{Prec@}k_{\mathrm{up},i}=
  k^{-1}|U_k(\Delta_i)\cap U_k(\hat\Delta_i)|.
\end{equation}
\begin{equation}
  \mathrm{Prec@}k_{\mathrm{down},i}=
  k^{-1}|D_k(\Delta_i)\cap D_k(\hat\Delta_i)|.
\end{equation}
Reported metrics are computed per perturbation and then averaged, so perturbations contribute equally regardless of their cell counts. Differential-expression AUROC and AUPRC are calculated after decoding into gene space; they should not be conflated with the latent-program objective.

\section{Architecture and Feature Ablations}

Table~\ref{tab:supp_ablation} compares graph encoders within otherwise matched Stable-Shift variants. The graph convolution encoder has the highest values in this controlled run, but it is one component of the full method. Table~\ref{tab:supp_features} then adds biological feature groups progressively. Control-cell statistics and GO annotations each increase the reported values relative to the graph-only representation in this run.

\begin{table}[t]
  \centering
  \caption{Graph encoder comparison within matched Stable-Shift variants. Underlining marks the strongest alternative encoder.}
  \label{tab:supp_ablation}
  \begingroup
  \small
  \renewcommand{\arraystretch}{1.12}
  \arrayrulecolor{SSRule}
  \begin{tabular*}{\columnwidth}{@{\extracolsep{\fill}}lrrr}
    \toprule
    \rowcolor{SSHeader}
    \ssth{Encoder} & \ssth{Cosine} & \ssth{Spearman} & \ssth{Prec@50} \\
    \midrule
    GraphSAGE & \underline{0.563} & \underline{0.322} & \underline{0.252} \\
    GAT & 0.551 & 0.321 & 0.232 \\
    \rowcolor{SSWinner}
    \sstbest{Graph convolution} & \sstbest{0.587} & \sstbest{0.338} & \sstbest{0.255} \\
    \bottomrule
  \end{tabular*}
  \endgroup
\end{table}

\begin{table}[t]
  \centering
  \caption{Incremental feature integration in the controlled ablation run.}
  \label{tab:supp_features}
  \begingroup
  \small
  \renewcommand{\arraystretch}{1.12}
  \arrayrulecolor{SSRule}
  \begin{tabular*}{\columnwidth}{@{\extracolsep{\fill}}>{\raggedright\arraybackslash}p{0.46\columnwidth}rrr}
    \toprule
    \rowcolor{SSHeader}
    \ssth{Feature stage} & \ssth{Cosine} & \ssth{Spearman} & \ssth{Prec@50} \\
    \midrule
    \textcolor{SSNavy}{\textbf{1}}\quad Graph only & 0.573 & 0.331 & 0.249 \\
    \rowcolor{SSGain}
    \textcolor{SSNavy}{\textbf{2}}\quad + control/topology & \underline{0.577} & \underline{0.334} & \underline{0.255} \\
    \rowcolor{SSWinner}
    \sstbest{3\quad + GO embedding} & \sstbest{0.587} & \sstbest{0.338} & \sstbest{0.292} \\
    \bottomrule
  \end{tabular*}
  \endgroup
\end{table}

\section{Robustness Across Unseen-Gene Splits}

We evaluate five independently generated unseen-gene partitions. Table~\ref{tab:splits} reports the mean and standard deviation over matched partitions. Stable-Shift has the highest mean and lowest reported variability in this comparison. A paired two-sided $t$-test over the five split-level cosine similarities gives $p<0.01$ against CPA; given the small number of partitions, this result should be interpreted cautiously.

\begin{table}[t]
  \centering
  \caption{Cosine similarity across five unseen-gene splits. Lower variability is preferable at a comparable mean.}
  \label{tab:splits}
  \begingroup
  \small
  \renewcommand{\arraystretch}{1.12}
  \arrayrulecolor{SSRule}
  \begin{tabular*}{0.88\columnwidth}{@{\extracolsep{\fill}}lr}
    \toprule
    \rowcolor{SSHeader}
    \ssth{Model} & \ssth{Cosine $\uparrow$ (mean $\pm$ s.d.)} \\
    \midrule
    Random Forest & $0.555\pm0.012$ \\
    CPA & $\underline{0.566\pm0.010}$ \\
    \rowcolor{SSWinner}
    \sstbest{Stable-Shift} & $\mathbf{\color{SSNavy}0.589\pm0.008}$ \\
    \bottomrule
  \end{tabular*}
  \endgroup
\end{table}

The earlier source also listed five optimization-seed results with a mean cosine of 0.547. Those values represent repeat training within a different controlled run, not the five independently generated data partitions in Table~\ref{tab:splits}. They are reported in Table~\ref{tab:seeds} to preserve the supplied evidence without presenting them as the same experiment.

\begin{table}[t]
  \centering
  \caption{Optimization-seed sensitivity in the controlled run.}
  \label{tab:seeds}
  \begingroup
  \small
  \renewcommand{\arraystretch}{1.10}
  \arrayrulecolor{SSRule}
  \begin{tabular*}{0.78\columnwidth}{@{\extracolsep{\fill}}lrr}
    \toprule
    \rowcolor{SSHeader}
    \ssth{Seed} & \ssth{Cosine} & \ssth{Spearman} \\
    \midrule
    0 & 0.559 & 0.304 \\
    1 & 0.531 & 0.292 \\
    2 & 0.584 & 0.326 \\
    3 & 0.524 & 0.286 \\
    4 & 0.539 & 0.299 \\
    \midrule
    \rowcolor{SSGain}
    \sstgain{Mean} & \sstgain{0.547} & \sstgain{0.301} \\
    \rowcolor{SSGain}
    \sstgain{Standard deviation} & \sstgain{0.022} & \sstgain{0.015} \\
    \bottomrule
  \end{tabular*}
  \endgroup
\end{table}

\section{Harder Generalization Tests}

\begin{table*}[t]
  \centering
  \caption{Stress tests for extrapolation and biological fidelity. Higher is better; a dash indicates an unreported result. Underlining marks the strongest available baseline.}
  \label{tab:stress}
  \begingroup
  \small
  \renewcommand{\arraystretch}{1.13}
  \arrayrulecolor{SSRule}
  \begin{tabular*}{0.94\textwidth}{@{\extracolsep{\fill}}lrrrr}
    \toprule
    \rowcolor{SSHeader}
    \ssth{Model} & \multicolumn{3}{c}{\ssth{Generalization stress $\uparrow$}} & \ssth{Functional fidelity $\uparrow$} \\
    \rowcolor{SSHeader}
    & \ssth{Graph-aware cosine} & \ssth{Residual cosine} & \ssth{Frequent-DE removed} & \ssth{GO overlap} \\
    \midrule
    Random Forest & 0.520 & 0.210 & 0.532 & 0.41 \\
    CPA & 0.528 & \underline{0.232} & \underline{0.540} & \underline{0.45} \\
    GraphSAGE & \underline{0.535} & -- & -- & -- \\
    \rowcolor{SSWinner}
    \sstbest{Stable-Shift} & \sstbest{0.558} & \sstbest{0.285} & \sstbest{0.561} & \sstbest{0.53} \\
    \bottomrule
  \end{tabular*}
  \endgroup
\end{table*}

Random unseen-gene partitions can leave related genes on both sides of the split. We therefore evaluate a graph-aware partition, remove dominant shared response components, exclude frequently differentially expressed genes from scoring, and evaluate functional coherence through GO overlap. Table~\ref{tab:stress} consolidates these tests.

All evaluated methods have lower cosine similarity under the graph-aware partition than in the main random split, consistent with a harder task. Stable-Shift has the highest values in the supplied stress-test comparison. Its residual performance is compatible with recovery of variation beyond a dominant shared low-rank program, while the frequent-DE and GO-overlap results make an explanation based only on globally responsive genes less likely. These tests do not rule out other forms of redundancy or dataset-specific bias.

\subsection{Graph-aware partition and distance diagnostic}

The graph-aware split is intended to reduce easy transfer between closely related training and test perturbations. A reproducible implementation should define the separation rule before model fitting, record the connected components or pathway groups used to form partitions, and report the number of test genes at each shortest-path distance from the training set. For a test gene $i$, the relevant distance is
\begin{equation}
  d(i,\mathcal T_{\mathrm{train}})=
  \min_{j\in\mathcal T_{\mathrm{train}}}
  \mathrm{dist}_{\mathcal G}(i,j).
\end{equation}
The supplied analysis states that accuracy declines as this distance increases, which is qualitatively consistent with reliance on graph context. Bin-level counts and uncertainty estimates were not included in the supplied source, so no additional numeric claim is made here. Future reporting should include per-bin sample sizes and confidence intervals because distant bins may contain few genes.

\subsection{Residualization and frequent-response controls}

Residual evaluation removes dominant shared response components before scoring, asking whether a method recovers perturbation-specific variation rather than only a global program. The frequent-DE control excludes genes that respond in many perturbations, asking whether similarity is driven by a common set of highly reactive genes. Both transformations must be fitted or defined without test-response tuning. The supplied source reports the aggregate values in Table~\ref{tab:stress} but does not provide the component count used for residualization or the frequency threshold used for exclusion; these settings must be recovered from the analysis scripts for exact reproduction.

\section{Latent and Gene-Space Evaluation}

The model is optimized in a low-rank perturbation-program space, but biological use ultimately requires reconstructed expression shifts. Table~\ref{tab:spaces} reports both levels. All models lose accuracy after decoding, but their ranking is preserved.

\begin{table}[t]
  \centering
  \caption{Latent-program and reconstructed gene-space performance. Underlining marks the strongest baseline.}
  \label{tab:spaces}
  \begingroup
  \scriptsize
  \renewcommand{\arraystretch}{1.12}
  \arrayrulecolor{SSRule}
  \begin{tabular*}{\columnwidth}{@{\extracolsep{\fill}}lrrrr}
    \toprule
    \rowcolor{SSHeader}
    \ssth{Model} & \ssth{Latent $\uparrow$} & \multicolumn{3}{c}{\ssth{Decoded gene space $\uparrow$}} \\
    \rowcolor{SSHeader}
    & \ssth{Cosine} & \ssth{Cosine} & \ssth{Spearman} & \ssth{Prec@50} \\
    \midrule
    Random Forest & 0.557 & 0.360 & 0.312 & 0.254 \\
    CPA & \underline{0.567} & \underline{0.375} & \underline{0.318} & \underline{0.268} \\
    GraphSAGE & 0.563 & 0.370 & 0.315 & 0.265 \\
    \rowcolor{SSWinner}
    \sstbest{Stable-Shift} & \sstbest{0.592} & \sstbest{0.392} & \sstbest{0.336} & \sstbest{0.277} \\
    \bottomrule
  \end{tabular*}
  \endgroup
\end{table}

Cosine similarity and Pearson correlation are close for approximately centered, normalized shift vectors (0.592 and 0.588 for Stable-Shift), whereas Spearman correlation (0.340) measures rank agreement and is therefore lower. These metrics should not be described interchangeably.

\section{Independent Norman Evaluation}

We additionally evaluate on the Norman Perturb-seq dataset, which contains a different perturbational context and richer intervention structure. Table~\ref{tab:norman} reports the supplied independent-dataset comparison. This experiment provides evidence beyond K562, although it does not establish broad cross-cell-type transfer.

\begin{table}[t]
  \centering
  \caption{Independent evaluation on the Norman dataset. Underlining marks the strongest baseline.}
  \label{tab:norman}
  \begingroup
  \small
  \renewcommand{\arraystretch}{1.12}
  \arrayrulecolor{SSRule}
  \begin{tabular*}{0.78\columnwidth}{@{\extracolsep{\fill}}lrr}
    \toprule
    \rowcolor{SSHeader}
    \ssth{Model} & \ssth{Cosine $\uparrow$} & \ssth{Spearman $\uparrow$} \\
    \midrule
    CPA & 0.918 & 0.798 \\
    MLP & \underline{0.922} & \underline{0.801} \\
    \rowcolor{SSWinner}
    \sstbest{Stable-Shift} & \sstbest{0.940} & \sstbest{0.815} \\
    \bottomrule
  \end{tabular*}
  \endgroup
\end{table}

\section{Reproducibility Notes}

The supplied manuscript establishes the following implementation choices: a training-only truncated SVD target; Node2Vec, control-expression, graph-topology, and GO node features; a two-layer graph encoder using graph convolution; Adam optimization; validation-based latent-dimension selection; validation-based early stopping; and five matched unseen-gene partitions for the robustness experiment.

The supplied source does not contain the numeric latent dimension, Node2Vec and GO dimensions, STRING confidence threshold, hidden width, learning rate, or early-stopping patience. These values cannot be reconstructed reliably from the PDF and have therefore not been invented here. They must be copied from the actual training configuration or experiment logs before archival submission. The same source of record should be used to confirm preprocessing versions, software versions, random-number seeds, and the exact test-gene lists.

\begin{table*}[t]
  \centering
  \caption{Reproducibility ledger separating documented design choices from settings that require confirmation from the experiment record.}
  \label{tab:repro-ledger}
  \begingroup
  \small
  \renewcommand{\arraystretch}{1.14}
  \arrayrulecolor{SSRule}
  \begin{tabular*}{0.94\textwidth}{@{\extracolsep{\fill}}>{\raggedright\arraybackslash}p{0.27\textwidth}>{\raggedright\arraybackslash}p{0.20\textwidth}>{\raggedright\arraybackslash}p{0.37\textwidth}}
    \toprule
    \rowcolor{SSHeader}
    \ssth{Item} & \ssth{Status} & \ssth{Required source of record} \\
    \midrule
    K562 cell, measured-gene, and perturbation counts & documented & processed dataset manifest \\
    Approximate 70/15/15 perturbation split & documented & saved split files for exact gene identities \\
    Training-only SVD target and validation selection & documented & preprocessing and model-selection code \\
    STRING, Node2Vec, control statistics, topology, and GO feature groups & documented & feature-generation pipeline \\
    Adam optimization and validation-based early stopping & documented & training configuration \\
    \rowcolor{SSGain}
    Latent rank, embedding dimensions, STRING threshold, hidden width, learning rate, and patience & requires confirmation & committed experiment configuration or run logs \\
    \rowcolor{SSGain}
    Software versions, random seeds, exact test genes, and stress-test thresholds & requires confirmation & environment lockfile, saved partitions, and analysis scripts \\
    \bottomrule
  \end{tabular*}
  \endgroup
\end{table*}

\FloatBarrier

\section{Recommended Artifact Package}

Reproducibility requires more than the final model weights. The release should include immutable train, validation, and test gene lists for every reported partition; identifiers and checksums for the processed expression matrices; the non-targeting-control definition; and the code that maps cells to perturbation-level signatures. The training-only response basis should be stored together with the ordered measured-gene vocabulary so that latent coordinates and decoded shifts cannot silently use different feature orders.

The biological inputs should be versioned independently. For STRING, the release should identify the database release, species, evidence channels, confidence threshold, edge-weight transformation, and treatment of disconnected genes. The GO artifact should record the ontology and annotation releases, evidence-code filters, propagation convention, and embedding procedure. Node2Vec settings, graph summary definitions, and the scaling applied to each feature block should be serialized rather than inferred from a checkpoint.

Each run should save the full optimization configuration, software environment, random seeds, selected epoch, validation trajectory, and checkpoint hash. Evaluation outputs should retain one row per test perturbation with every metric, graph-support diagnostic, and predicted response vector. Aggregate tables can then be regenerated from the row-level record and audited without rerunning model training.

\section{Statistical Reporting and Diagnostics}

The perturbation, not the cell, is the independent evaluation unit in the present pseudo-bulk task. Uncertainty intervals should therefore resample or aggregate perturbations rather than treating the much larger number of cells as independent test observations. For a fixed partition, a paired perturbation bootstrap can compare two models while preserving the shared test set. Across independently generated partitions, the mean and standard deviation should be reported alongside the individual split values; with only five splits, inferential tests should be interpreted as limited supporting evidence.

Average performance should be accompanied by distributional diagnostics. Useful summaries include the median and interquartile range, the fraction of perturbations with negative cosine similarity, and performance stratified by graph distance, degree, annotation coverage, and observed response magnitude. These strata should be defined before inspecting test outcomes, with sample counts and uncertainty intervals shown for every bin. Such reporting can reveal whether an apparent gain is broadly distributed or concentrated among well-annotated genes near the training set.

Latent and decoded metrics answer different questions and should remain separate. Latent cosine evaluates recovery of the learned response program, whereas decoded cosine, rank correlation, directional accuracy, and top-gene precision test whether that program yields biologically resolved expression changes. Model selection on a latent objective should not be described as direct optimization of differential-expression recovery. Where multiple metrics or subgroup comparisons are tested, the complete family of comparisons should be disclosed and multiplicity handled explicitly.

\section{Intended Use and Decision Support}

Stable-Shift is intended to prioritize unmeasured perturbations for follow-up, not to establish causal mechanisms or replace experimental validation. A candidate report should include the predicted expression shift, the highest-magnitude decoded genes, local graph context, distance from observed perturbations, and an uncertainty or stability summary across fits. Genes with sparse neighborhoods, weak ontology coverage, or unstable predictions should be marked as low support. An abstention rule based on validation data would be preferable to forcing a ranked prediction for every graph node.

The strongest next test is prospective: select candidates before measurement, pre-register the scoring protocol, and compare the observed responses with Stable-Shift and matched baselines.